%% file: SM_update_Primary_End.tex
\pdfoutput=1
\documentclass[aps, prl, reprint, superscriptaddress, nofootinbib]{revtex4-1}
\usepackage{amsmath, amsfonts, amssymb, hyperref, color, graphicx, xspace, enumitem, slashed}
\usepackage[T1]{fontenc}
\definecolor{nicered}{rgb}{.7,.1,.1}
\definecolor{nicegreen}{rgb}{.1,.5,.1}
\definecolor{darkblue}{rgb}{0,0,.5}
\hypersetup{colorlinks, citecolor=nicegreen, linkcolor=nicered, urlcolor=darkblue}

\DeclareMathOperator{\rank}{Rank}

\usepackage{tikz}
\usetikzlibrary{arrows,shapes}
\usetikzlibrary{trees}
\usetikzlibrary{matrix,arrows} 				
\usetikzlibrary{calc} 						
\usetikzlibrary{positioning}					
\usetikzlibrary{calc,through}				
\usetikzlibrary{decorations.pathreplacing}  	
\usepackage[tikz]{bclogo} 					
\usepackage{pgffor}						
\usetikzlibrary{decorations.pathmorphing}		
\usetikzlibrary{decorations.markings}
\usetikzlibrary{intersections}				

\begin{document}

\title{Revising the full one-loop gauge prefactor in electroweak vacuum stability}

\author{Pietro Baratella}
\email{pietro.baratella@ijs.si}
\affiliation{Jo\v{z}ef Stefan Institute, Jamova 39, 1000 Ljubljana, Slovenia}

\author{Miha Nemev\v{s}ek}
\email{miha.nemevsek@ijs.si}
\affiliation{Jo\v{z}ef Stefan Institute, Jamova 39, 1000 Ljubljana, Slovenia}
\affiliation{Faculty of Mathematics and Physics, University of Ljubljana, Jadranska 19, 
1000 Ljubljana, Slovenia}

\author{Yutaro Shoji}
\email{yutaro.shoji@ijs.si}
\affiliation{Jo\v{z}ef Stefan Institute, Jamova 39, 1000 Ljubljana, Slovenia}

\author{Katarina Trailović}
\email{katarina.trailovic@ijs.si}
\affiliation{Jo\v{z}ef Stefan Institute, Jamova 39, 1000 Ljubljana, Slovenia}
\affiliation{Faculty of Mathematics and Physics, University of Ljubljana, Jadranska 19, 
1000 Ljubljana, Slovenia}

\author{Lorenzo Ubaldi}
\email{lorenzo.ubaldi@ijs.si}
\affiliation{Jo\v{z}ef Stefan Institute, Jamova 39, 1000 Ljubljana, Slovenia}

\date{\today}

\begin{abstract}
We revisit the decay rate of the electroweak vacuum in the Standard Model with the full one-loop prefactor.
We focus on the gauge degrees of freedom and derive the degeneracy factors appearing in the functional 
determinant using group theoretical arguments.
Our treatment shows that the transverse modes were previously overcounted, so we revise the calculation 
of that part of the prefactor.
The new result modifies the gauge fields' contribution by $6\%$ and slightly decreases the previously predicted 
lifetime of the electroweak vacuum, which remains much longer than the age of the universe.
Our discussion of the transverse mode degeneracy applies to any calculation of functional determinants 
involving gauge fields in four dimensions.
\end{abstract}

\maketitle

{\bf \textsc{Introduction.}}
The Standard Model of particle physics (SM) is the cornerstone of our understanding of elementary particle 
interactions. 
The only fundamental scalar field in the SM is the Higgs doublet $H$, responsible for the spontaneous breaking 
of the $SU(2)_L \times U(1)_Y$ electroweak (EW) gauge symmetry.
With the current central values of SM parameters, the potential $V(h)$ of the physical Higgs $h$, once 
extrapolated to the regime of an extremely intense field, turns negative for 
$\langle h \rangle \gtrsim 10^{10} \text{ GeV} \gg v \approx 246 \text{ GeV}$.
This makes the EW vacuum at $\langle h \rangle = v$ metastable due to quantum tunneling.

The tunneling rate per unit volume $\gamma$ can be computed using the methods 
of~\cite{Coleman:1977py, Callan:1977pt}, and expressed as $\gamma = {\cal A} \, e^{-S}$, where $S$ is the 
action of the bounce in Euclidean spacetime and ${\cal A}$ is the prefactor with mass dimension four.
The bounce $\overline{h}(\rho)$ is an $O(4)$-symmetric instanton solution in terms of the Euclidean radius 
$\rho^2 = t^2 + |{\bf x}|^2$ that connects a point close to the absolute vacuum at $\rho = 0$ to the unstable 
vacuum at $\rho = \infty$.
It is computed by approximating the Higgs potential as $\lambda (H^\dagger H)^2$, with $\lambda$ negative, 
thus neglecting the quadratic term and treating $v \sim 0$.
This is justified because the Higgs field travels over large field values until the potential becomes negative.
In this approximation the potential is classically scale invariant and the bounce is given by the Fubini-Lipatov 
instanton $\overline h(\rho) = \sqrt{8 / |\lambda |} [R/(\rho^2 + R^2)]$, whose action is $S = 8\pi^2 / (3|\lambda |)$.
The free parameter $R$ signals the classical symmetry under dilatations, which is broken by quantum corrections.
This effectively fixes $R^{-1} \sim 10^{17}$ GeV, which is the scale where the beta function of the running quartic 
coupling $\lambda$ vanishes, assuming only SM degrees of freedom. 

To obtain ${\cal A}$ one has to compute the functional determinants corresponding to one-loop diagrams,
where the fields running in the loop are the scalar, fermion, and gauge boson fluctuations that couple to the 
Higgs bounce $\overline h(\rho)$.
Collecting the dominant contributions we can write
\begin{equation} \label{eq:totprefactor}
  {\cal A} = V_G\, {\cal A}^{(h)}{\cal A}^{(t)}{\cal A}^{(Z,\varphi_Z)}{\cal A}^{(W^\pm,\varphi^\pm)} \, .
\end{equation}
Here, $V_G = 2 \pi^2$ is the volume of the $SU(2)$ group broken by the bounce, the superscripts denote 
the species: $t$ is the top quark, $Z$ and $W^\pm$ the gauge bosons and $\varphi_Z, \varphi^\pm$ the 
would-be Nambu-Goldstone bosons (NGBs).
We ignore the dilatational zero mode in~\eqref{eq:totprefactor} for the moment and come back to it later.

\begin{figure}
  \vspace{.4cm}
  \centering
  \includegraphics[width=\columnwidth]{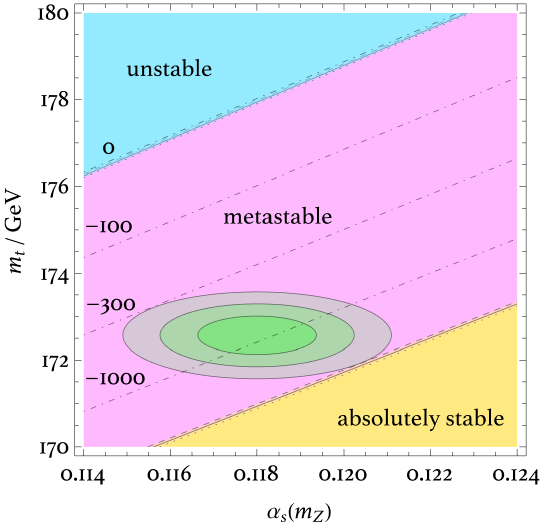}
  \caption{
  Contours of $\gamma$ as a function of $\alpha_s(m_Z)$ and the top mass $m_t$.
  The black dot-dashed contours correspond to $\gamma = 1, 10^{-100}, 10^{-300}$ and 
  $10^{-1000} \text{ Gyr}^{-1} \text{Gpc}^{-3}$ for the Higgs mass $m_h =125.20$ GeV. 
  In the blue region $\gamma$ becomes larger than $H_0^4$, with $H_0$ the current Hubble parameter,
  while in the yellow region the EW vacuum is stable.
  The boundaries of these regions are given by the plain lines for $m_h =125.20$ GeV, the dotted 
  lines for $m_h =125.09$ GeV, the dashed line for $m_h =125.31$ GeV.
  The green region, for which $\gamma = 10^{-871}  \text{ Gyr}^{-1} \text{Gpc}^{-3}$ at the center, shows the 
  experimentally measured values of $\alpha_s(m_Z)$ and the top mass with their $1\sigma$ (inside), $2\sigma$ 
  (middle) and $3\sigma$ (outside) uncertainties in quadrature.
  \vspace{-4ex}}
  \label{fig:SM_FV}
\end{figure}

The stability of the EW vacuum has been investigated by several authors~\cite{Sher:1988mj, Casas:1994qy, 
Espinosa:1995se, Isidori:2001bm, Espinosa:2007qp, Ellis:2009tp, Degrassi:2012ry, Buttazzo:2013uya,
Lalak:2014qua, Andreassen:2014gha, Bednyakov:2015sca, Branchina:2014rva, Iacobellis:2016eof}.
The calculation of the full one-loop prefactor was first done in~\cite{Isidori:2001bm} before the Higgs discovery 
and updated in~\cite{Andreassen:2017rzq, Chigusa:2017dux, Chigusa:2018uuj} with a careful treatment of 
dilatation~\cite{Andreassen:2017rzq,Chigusa:2017dux,Chigusa:2018uuj} and gauge zero 
modes~\cite{Endo:2017gal, Endo:2017tsz}.

In this Letter we revisit the calculation of the gauge prefactor.
We find that the transverse mode degeneracy was not properly taken into account.
Once corrected, the central value of the SM rate $\gamma$ increases only slightly, from 
$10^{-877}$ to $10^{-871}$ Gyr$^{-1}$ Gpc$^{-3}$.
The SM vacuum lifetime remains longer than the current age of the universe and there is no 
occasion for anxiety~\cite{Coleman:1977py}.

We first introduce the fluctuation operator in the gauge sector.
Then we explain how to build suitable bases for the scalar and gauge fields, given the Euclidean spherical 
$4D$ symmetry of the bounce, and discuss the counting of degeneracy factors.
Next, we obtain the analytic expression of our correction to the vacuum decay rate in the SM and give the 
numerical full one-loop vacuum decay rate with the current central values of the SM couplings. 
Our final result is summarized in FIG.~\ref{fig:SM_FV}. 

{\bf \textsc{Gauge fluctuation operator.}}
In the presence of the bounce, gauge fields and NGBs mix.
The fluctuation operator does not mix $Z_\mu$ and $W_\mu^\pm$ in the SM, so we unify the notation as 
$A_\mu = Z_\mu,W_\mu^\pm$ and $\varphi = \varphi_Z, \varphi^\pm$.
The prefactor, coming from $A_\mu$ and $\varphi$, is then
\begin{align} \label{eq:det_ratio}
  {\cal A}^{(A, \varphi)} = J_G \left( \frac{\det' S''^{(A, \varphi)}}{\det \hat S''^{(A, \varphi)}} \right)^{-\frac{1}{2}} \, ,
\end{align}
where the prime indicates the subtraction of the zero mode, associated with the spontaneous breaking of gauge 
symmetry, and $J_G$ is the group space integral Jacobian.
The fluctuation operator in the $(A^\mu,\varphi)$ basis is given by the $5\times 5$ matrix
\begin{align} \label{eq:fluct_operator}
  \!\!\! S''^{(A,\varphi)} = \begin{pmatrix}
  \left(- \partial^2 +\frac{1}{4}g^2\overline{h}{ }^2 \right)\delta_{\mu\nu} & \frac{1}{2}g \overline{h}' \hat x_\nu-
  \frac{1}{2}g \overline{h} \partial_\nu 
  \\
  g \overline{h}'\hat x_\mu+\frac{1}{2}g \overline{h}\partial_\mu & -\partial^2 + \lambda \overline{h}{ }^2
  \end{pmatrix}\ ,
\end{align}
while $\hat S''^{(A, \varphi)}$ is the same operator, but with $\bar h = 0$.
Here, $\bar h'=\partial_\rho \bar h$, $g$ is the gauge coupling for $W^{\pm}$ or $Z$, and $\hat{x}_\mu$ is a unit 
vector, such that $x_\mu = \rho \, \hat{x}_\mu$.
We work in the Fermi gauge with $\xi = 1$, which is defined through the gauge fixing term 
$\mathcal L_{\rm GF} = (\partial_\mu A^\mu)^2/2$.
Given the $O(4)$ symmetry of the bounce, the 4$D$ Laplacian is conveniently written in spherical coordinates as
$\partial^2 = \partial_\rho^2 + 3 \rho^{-1} \partial_\rho - L^2 / \rho^2$, where $L^2 = L^{\mu\nu}L_{\mu\nu}/2$,
and $L_{\mu\nu} = -i (x_\mu\partial_\nu - x_\nu \partial_\mu)$ is the orbital angular momentum operator.

The fluctuation operator commutes with rotations, and therefore with all the components of the total
angular momentum operator, $[J_{\mu\nu}, S''^{(A,\varphi)}] = 0$.
This implies that only modes with the same total angular momentum quantum numbers mix under
the action of \eqref{eq:fluct_operator}.
In the following we decompose $\varphi$ and $A_\mu$ into $J_{\mu\nu}$ bases.
In the calculation of functional determinants, it is the {\em total}, not the orbital, angular momentum operator 
that dictates the counting of degeneracies.

{\bf \textsc{Scalar field basis.}}
The scalar field transforms under Euclidean rotations like $\varphi(x) \to \varphi(R^{-1}x)$,
where $R$ is a finite rotation of coordinates.
For the scalar, the total and orbital angular momenta coincide, $J_{\mu\nu} = L_{\mu\nu}$.
The space of scalar fields carries an infinite dimensional unitary representation of the compact group $SO(4)$ 
that, thanks to the Peter-Weyl theorem, admits an orthogonal decomposition into irreducible finite dimensional 
representations (irreps).
$SO(4)$ is locally isomorphic to $SU(2)_A \otimes SU(2)_B$.
Following textbook conventions, in $SU(2)_A$ we define the total angular momentum operator in the $\hat 3$ 
direction as $J_{A3}$ and the Casimir operator as $J_A^2$.
The latter has eigenvalues $j_A(j_A + 1)$ with $j_A$ a half-integer.
The same holds for $SU(2)_B$ and we label the irreps with $(j_A, j_B)$, with dimension of $(2j_A+1)(2j_B+1)$. 

The representation space of the scalar field splits into $\oplus_{j = 0}^\infty (j/2, j/2)$, where the label 
$j = j_A + j_B$ is an integer.
Each $(j/2,j/2)$ multiplet is an eigenstate of $J^{\mu\nu}J_{\mu\nu}/2 = J^2 = 2(J_A^2 + J_B^2)$ 
with eigenvalue $j(j+2)$ and degeneracy factor
$ d_j^{(\varphi)} \equiv {\rm dim} \left( j/2, j/2 \right) = \left( j + 1 \right)^2$.

An explicit basis for such irreps can be formed with hyperspherical harmonics $Y_{j m_A m_B}$
that satisfy $L^2 Y_{j m_A m_B} = j (j + 2) Y_{j m_A m_B}$.
Here, $m_{A}$ and $m_{B}$ run in integer steps between $-j/2$ and $j/2$, giving the multiplicity $d_j^{(\varphi)}$.
The NGB $\varphi$ is written as
\begin{align}\label{eq:SH}
  \varphi(x)=\sum_{j=0}^\infty\sum_{m_{A}, m_{B}} \varphi_{j m_{A}m_{B}}(\rho)\ Y_{j m_{A} m_{B}}(\hat{x}) \, .
\end{align}

{\bf \textsc{Vector field basis.}}
For the spin-1 field, the total angular momentum is $J_{\mu \nu} = L_{\mu\nu} + S_{\mu \nu}$, where 
$[S_{\mu \nu}]_{\rho\sigma} = -i( \delta_{\mu\rho}\delta_{\nu\sigma}-\delta_{\mu\sigma}\delta_{\nu\rho})$ are 
the generators acting on the spin-component. 
Equivalently, the vector field transforms like $A_\mu(x) \to R_{\mu \nu} \, A_\nu(R^{-1} x)$, where $R_{\mu \nu}$ 
is the vector rotation matrix.
The representation of $A_\mu(x)$ is isomorphic to the tensor product of a spin-1 component $(1/2, 1/2)$, 
with an orbital component, transforming as $\oplus_{l = 0}^\infty(l/2, l/2)$.
To understand the $SO(4)$ irreps $(j_A, j_B)$ of the vector field, we expand the tensor product:
\begin{align}   \label{eq:MultiL}
\begin{split}
  &\left(\frac12, \frac12 \right) \ \otimes \ \bigoplus_{l=0}^\infty \ \left(\frac{l}{2}, \frac{l}{2}\right) =  
  \\
  & \hspace{1ex}  \left(\frac12,\frac12 \right) \oplus \bigoplus_{l = 1}^\infty \, \left[
  \left( \frac{l+1}{2}, \frac{l+1}{2} \right) \oplus \left( \frac{l -1}{2}, \frac{l - 1}{2} \right) \right.
  \\
  & \hspace{1ex} \left. \oplus 
  \left( \frac{l+1}{2}, \frac{l-1}{2}  \right) \oplus \left( \frac{l-1}{2}, \frac{l+1}{2}  \right) \, \right]\, = 
\end{split}
\\
\begin{split}
  &(0, 0)_{l=1} \oplus \, \bigoplus_{j = 1}^\infty \, \left[\left( \frac{j}{2}, \frac{j}{2}  \right)_{l = j + 1}
  \oplus\left( \frac{j}{2}, \frac{j}{2} \right)_{l = j - 1} \right] \oplus 
  \\
  & \hspace{2ex} 
  \bigoplus_{j = 1}^\infty  \, \left[ \left( \frac{j+1}{2}, \frac{j-1}{2}  \right)_{l = j}\oplus \left( \frac{j-1}{2}, 
  \frac{j+1}{2} \right)_{l = j} \, \right] \, .
  \end{split}
  \label{eq:MultiJ}
\end{align}
\begin{figure}[t]
  \centering
  \input{MultipoleScheme.tikz}
  \vspace{.5cm}
 \caption{
   Decomposition of a 4-vector field $A_\mu(\hat{x})$ under $SO(4)$.
   The diagonal modes ${\cal D}_j^\pm$ with $j_A = j_B$ appear in double copies, except for $(0,0)$.
   The off-diagonal multiplets with $j_A = j_B \pm 1$ correspond to the transverse modes ${\cal T}_j^{\pm}$, 
   and appear as single copies.
   Blue circles with $l = 3$ exemplify how the $L^2$ eigenspace gets distributed within the $(j_A, j_B)$ lattice.
   \vspace{-4ex}} \label{fig:irreps_A}
\end{figure}
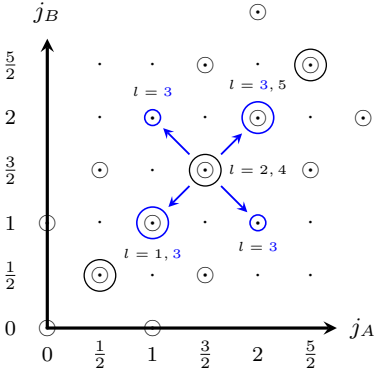
To go from \eqref{eq:MultiL} to \eqref{eq:MultiJ} we simply re-labeled and rearranged the sum.
In \eqref{eq:MultiJ} we included a subscript for each $(j_A,j_B)$ multiplet to track the orbital angular momentum 
quantum number $l$.
We call the multiplets with $j_A = j_B$ the diagonal modes ${\cal D}^\pm_j$, corresponding to $(j/2, j/2)_{j \pm 1}$
(note ${\cal D}^-_0$ is zero); they have eigenvalue $j(j+2)$ under $J^2$, and degeneracy 
$d_j^{\cal D}  \equiv {\rm dim} \left( \frac{j}{2}, \frac{j}{2} \right) = \left( j + 1 \right)^2$.
These are the same total angular momentum quantum numbers as the NGB scalars, hence in general the fluctuation 
operator matrix mixes these states.
We identify the multiplets having $j_A = j_B \pm 1$ with the transverse modes~\cite{Higuchi:1991tn}, 
and call them ${\cal T}^\pm_j$ for $((j\pm1)/2,(j \mp1)/2)_j$.
Their $J^2$ eigenvalue is $(j+1)^2$, while the degeneracy factor is
\begin{align}
 \label{eq:djT}
  d_j^{\cal T}	& \equiv {\rm dim} \left( \frac{j \pm 1}{2}, \frac{j \mp 1}{2} \right) = j \left( j + 2 \right) \, .
\end{align}
FIG.~\ref{fig:irreps_A} summarizes the group theoretical construction and there ${\cal D}^\pm_j$ correspond to 
the diagonal circles, where the two overlapping circles are distinguished by the eigenvalues of $L^2$.
The ${\cal T}^\pm_j$ correspond to the lower and upper off-diagonal circles, respectively.

We decompose $A_\mu(x)$ as a sum of radial functions times a basis of vector fields,
\begin{align} \label{eq:AmuBasis}
  \!\!\!\!\!\!A_\mu(x) = \sum_{j_A, j_B, l, m_A, m_B} A^{j_A j_B m_A m_B}_l(\rho) V^{j_A j_B m_A m_B}_{l;\mu}(\hat x) \, ,
\end{align}
where the real field $A_\mu$ is expanded with a complex basis, without affecting the degeneracy, as follows
\begin{align} \label{eq:Vlmu}
\begin{split}
  &V^{j_A j_B m_A m_B}_{l;\mu}(\hat x) = \sum_{m_{l_A},m_{l_B}, m_{s_A},m_{s_B}}
  C^{j_A m_A}_{\frac{l}{2} m_{l_A} \frac{1}{2} m_{s_A}}
  \\
  &\hspace{2ex}  C^{j_B m_B}_{\frac{l}{2} m_{l_B} \frac{1}{2} m_{s_B}}  
  \left( \tilde \sigma_\mu \right)_{m_{s_A} m_{s_B}} \, 
  Y_{l m_{l_A} m_{l_B}}(\hat x) \, .
\end{split}
\end{align}
This choice has the virtue of containing familiar objects that are in direct correspondence with the group 
theory construction of FIG.~\ref{fig:irreps_A}.
The $(\tilde \sigma_\mu)_{m_{s_A} m_{s_B}}$ matrices correspond to the $(1/2,1/2)$ object that ensures 
the proper transformation under rotations, and are given by $\tilde \sigma_\mu =  \varepsilon \cdot \sigma_\mu$,
where $\varepsilon$ is the two-dimensional Levi-Civita symbol, $\sigma_{1,2,3}$ are the usual Pauli matrices, 
and $\sigma_4 = \text{diag}(i ,i )$. 
The entries of $\tilde{\sigma}$ are ordered such that $m_s=1/2$ comes before $m_s=-1/2$.
$C^{j_A,m_A}_{l_A m_{l_A} s_A m_{s_A}}$ denote the $SU(2)_A$ Clebsch-Gordan coefficients (and the 
same for $B$).
These are non-zero only when $| 2 j_{A, B}-l | = 1$, $m_A = m_{l_A} + m_{s_A}$ and 
$m_B = m_{l_B} + m_{s_B}$.
We shall suppress the $m$ indices from here on for brevity; they label states with the same fluctuation 
operator and their summation only appears as the degeneracy factor.
Then, the non-zero components of $V^{j_A j_B}_{l;\mu}$ are $V^{\frac{j}{2} \frac{j}{2}}_{l=j\pm1;\mu}$ and 
$V^{\frac{j\pm1}{2} \frac{j\mp1}{2}}_{l = j;\mu}$, corresponding to the basis of 
${\cal D}_j^\pm$ and ${\cal T}_j^\pm$.
These components are eigenfunctions of $J^2$, with the eigenvalues quoted earlier for 
${\cal D}_j^\pm$ and ${\cal T}_j^\pm$.  
Each component is an eigenfunction of $L^2$ that acts only on $Y_{l m_{l_A} m_{l_B}}$.

{\bf \textsc{Fluctuation operator decomposition.}}
Let us decompose~\eqref{eq:fluct_operator} using the scalar and gauge basis functions in~\eqref{eq:SH} 
and~\eqref{eq:AmuBasis}.
When~\eqref{eq:fluct_operator} acts on $(A_\mu, \varphi)$, it splits into an infinite number of blocks with the 
same $j_A, j_B, m_A$ and $m_B$: $(A_1^{00},\varphi_0)$ for $j=0$, and $(A_{j+1}^{\frac{j}{2}\frac{j}{2}},A_{j-1}^{
\frac{j}{2}\frac{j}{2}},\varphi_j)$, $(A_{j}^{\frac{j+1}{2}\frac{j-1}{2}})$, $(A_{j}^{\frac{j-1}{2}\frac{j+1}{2}})$ for $j > 0$.
The first two indicate $2 \times 2$ or $3 \times 3$ fluctuation matrices that mix the ${\cal D}^\pm_j$ modes 
with the NGB, and the last two are those for the ${\cal T}^\pm_j$ modes that do not mix.

With this decomposition, the prefactor in~\eqref{eq:det_ratio} factorizes into 
${\cal A}^{(A,\varphi)} = {\cal A}^{({\cal D}, \varphi)} {\cal A}^{({\cal T})}$.
The first term ${\cal A}^{({\cal D} , \varphi)}$ was already computed correctly in the previous 
literature~\cite{Isidori:2001bm, Andreassen:2017rzq, Chigusa:2017dux}.
Therefore we only deal with the second one:
\begin{equation}  \label{eq:A_transverse}
  {\cal A}^{({\cal T})} = \left[ \prod_{j = 1}^\infty \left( \frac{\det S''^{({\cal T})}_j}{\det \hat S''^{({\cal T})}_j} 
  \right)^{2 d_j^{\cal T}} \right]^{-\frac{1}{2}} \, .
\end{equation}
The factor of $2$ in the exponent $2 d_j^{\cal T}$ comes from the two transverse modes ${\cal T}^\pm_j$ 
having the same fluctuation operator, given by $S''^{({\cal T})}_j = -\Delta_j + g^2 \overline{h}{}^2/4$, where 
$\Delta_{j = l} = \partial_\rho^2 + 3 \rho^{-1} \partial_\rho - l(l + 2) \rho^{-2}$.
There are no zero modes to worry about in~\eqref{eq:A_transverse}, so there is no prime in the 
determinant at the numerator.

\begin{itemize}
\item In previous calculations~\cite{Isidori:2001bm, Andreassen:2017rzq, Chigusa:2017dux}, the degeneracy 
factor for the transverse modes ${\cal T}_j^\pm$ was erroneously set equal to $(j+1)^2$.
The correct one, as we have shown in \eqref{eq:djT}, is $d_j^{\cal T} = j(j+2)$.
This leads to a slightly different result for the prefactor ${\cal A}^{(\cal T)}$.
We will shortly revise that calculation. 
\item For the diagonal
modes ${\cal D}^\pm_j$, our basis functions have one-to-one correspondence to the
ones given in~\cite{Isidori:2001bm}, which are then used 
in~\cite{Andreassen:2017rzq, Chigusa:2017dux, Chigusa:2018uuj}.
We differ for the transverse modes ${\cal T}^\pm_j$.
In the appendix of the end matter we give an explicit demonstration of why the gauge boson basis in~\cite{Isidori:2001bm}
has inconsistencies.
In the supplemental material we furhter elaborate and show that their basis functions are linearly dependent and do 
not span the entire space.
Despite the degeneracy factor and completeness issues, the operator $S''^{({\cal T})}_j$ 
in~\cite{Isidori:2001bm, Andreassen:2017rzq, Chigusa:2017dux} is correct.
\item Ref.~\cite{Isidori:2001bm} stated that, if one works in the background gauge with $\xi = 1$, {\it i.e.} 
$\mathcal L_{\rm GF}=(\partial_\mu A^\mu-g\bar h\varphi/2)^2/2$, the transverse and ghost modes mutually cancel.
This is incorrect, given the corrected $d_j^{\cal T}$ in \eqref{eq:djT}; one does need to include the ghost modes
in this gauge.
\end{itemize}

{\bf \textsc{Functional determinants for ${\cal T}^\pm$-modes.}}
The infinite product in ${\cal A}^{({\cal T})}$ is ultraviolet divergent.
A commonly employed regularization method subtracts the $\mathcal O(\delta S'',\delta S''^2)$ terms 
from $\ln {\cal A}^{(A,\varphi)}$, and adds back their dimensionally regularized quantities \cite{Isidori:2001bm}.
Here, $\delta S'' = S''^{(A \varphi)}-\hat S''^{(A \varphi)}$.
The added-back terms do not suffer from the mode counting subtlety since they are calculated in 
momentum space.
We focus on the subtraction procedure and defer to existing literature for the rest.

The finite part with the correct degeneracy factor is
\begin{equation}
  \!\!\!\! \ln {\cal A}^{({\cal T})}_{\rm fin} = -\sum_{j = 1}^\infty d_j^{\cal T} \left( \ln R_j - \ln R_j^{(1)} - \ln R_j^{(2)} \right) \, ,
\end{equation}
and it was finite even with the incorrect $d_j^{\cal T}$.
The analytic formula~\cite{Andreassen:2017rzq, Chigusa:2018uuj} for each term is
\begin{align}
  \ln R_j \equiv \ln\frac{\det S''^{({\cal T})}_j}{\det \hat S''^{({\cal T})}_j} &= \ln 
  \frac{\Gamma(j + 1) \Gamma(j + 2)}{\Gamma(j + \frac{3 + \kappa}{2}) \Gamma(j + \frac{3 - \kappa}{2})} \,,
\end{align}
and
\begin{align}
  \ln R_j^{(1)} &\equiv \left[\ln\frac{\det S''^{({\cal T})}_j}{\det \hat S''^{({\cal T})}_j}\right]_{\mathcal O(\delta S'')} =
  \frac{g^2}{2 |\lambda|} \frac{1}{j+1} \,,
  \\
  \begin{split}
  \ln R_j^{(2)} &\equiv \left[\ln\frac{ \det S''^{({\cal T})}_j}{\det \hat S''^{({\cal T})}_j}\right]_{\mathcal O(\delta S''^2)}
  \\
  &= \frac{g^4}{4\lambda^2} \left(\frac{2j+3}{2(j+1)^2}-\psi'(j+1)\right) \, ,
  \end{split}
\end{align}
with $\kappa = \sqrt{1 - 2 g^2/|\lambda|}$ and $\psi(z)$ the digamma function.

As all the other fluctutations in the prefactor \eqref{eq:totprefactor} were already computed correctly, we define
the correction as
\begin{equation}
 \delta \ln {\cal A}  = \ln {\cal A}^{(A,\varphi)}_{\rm prev}- \ln {\cal A}^{(A,\varphi)},
\end{equation}
where $\ln {\cal A}^{(A,\varphi)}_{\rm prev}$ is the gauge prefactor in~\cite{Andreassen:2017rzq, Chigusa:2018uuj}, 
and $\ln {\cal A}^{(A,\varphi)}$ the one computed with the correct degeneracy factor.
We find
\begin{align}
\begin{split}
  &\delta \ln {\cal A}=1-\frac{g^4}{16|\lambda|^2}(10-\pi^2)-\frac{g^2}{2|\lambda|}(2-\gamma)
  \\
  &\hspace{2ex}-\frac52\ln2\pi+\ln\left(\frac{4|\lambda|}{g^2} \cos \left(\frac{\pi}{2}\kappa \right) \right)
  \\
  &\hspace{2ex}-\frac{1-\kappa}{2}\ln\Gamma\left(\frac{3-\kappa}{2}\right)-\frac{1+\kappa}{2}\ln\Gamma\left(\frac{3+\kappa}{2}\right)
  \\
  &\hspace{2ex}+\psi^{(-2)}\left(\frac{3-\kappa}{2}\right)+\psi^{(-2)}\left(\frac{3+\kappa}{2}\right) \, .
\end{split}
\end{align}

{\bf \textsc{Vacuum decay rate.}}
To compute the final updated decay rate, we rely on the analytic formul\ae\  for all the remaining prefactors 
calculated in~\cite{Andreassen:2017rzq,Chigusa:2018uuj}.
We follow~\cite{Chigusa:2017dux, Chigusa:2018uuj} for the evaluation of the SM couplings at high energy.
The gauge, top Yukawa, and Higgs quartic couplings at $\mu = m_t$ are determined using the fitting formul\ae\
in~\cite{Buttazzo:2013uya}, at NNLO precision.
The bottom and tau Yukawa couplings are determined by $y_b(m_b) = \sqrt{2} m_b(m_b)/v$ and 
$y_\tau(m_\tau)=\sqrt{2} m_\tau/v$.
We ignore threshold corrections to bottom and tau masses; their effect on the vacuum decay rate 
is negligible.
We run these couplings using the three-loop $\beta$-functions given in~\cite{Buttazzo:2013uya}. 
The calculation of the rate involves an integral over the dilatation parameter $R$.
The dilatation symmetry is broken by the running of the couplings and we include this effect by taking the 
renormalization scale to $\mu = 1/R$, following~\cite{Chigusa:2017dux, Chigusa:2018uuj}.
We also put a UV cut-off to the integral such that $1/R$ or the maximum value of the Higgs field do not exceed 
the Planck scale~\cite{Chigusa:2018uuj}.
The rate is then calculated using the public code {\tt ELVAS}~\cite{Chigusa:2017dux,Chigusa:2018uuj}, with 
the modification of the transverse mode degeneracy in the gauge sector. 

We use the current values of the SM parameters and their $1\sigma$ errors from~\cite{ParticleDataGroup:2024cfk}: 
$m_h = 125.20 \pm 0.11 \text{ GeV}$, $m_t = 172.57 \pm 0.29 \text{ GeV}$ and $\alpha_s(m_Z) = 0.1180 \pm 0.0009$.
Our final decay rate is
\begin{equation}\label{eq:sm_result}
  \log_{10}\frac{\gamma}{\rm Gyr^{-1} \, Gpc^{-3}} = -871^{+35+175+209}_{-37-253-330} \, ,
\end{equation}
where the first, second, and third errors are calculated by changing the Higgs mass, the top mass, and the 
strong coupling by $1\sigma$, respectively.
The variations are summarized in~FIG.~\ref{fig:SM_FV}. 

Comparing to the most recent previous result~\cite{Chigusa:2017dux}, we find 
$\gamma / \gamma_{\rm prev} \approx 10^6$, using the central values of the SM parameters. 
The change of the central value is much smaller than the uncertainties in~\eqref{eq:sm_result}. 
However, it is still larger than the theoretical uncertainty ($\approx 10^3$), evaluated by setting the 
renormalization scale to $\mu = 2/R$ and $\mu = 1/(2R)$ instead of $\mu = 1/R$.

{\bf \textsc{Summary.}}
We revisited the total angular momentum decomposition of gauge fields in $4D$ and found an overcounting of 
degeneracies of the transverse modes gauge fluctuations in previous literature.
We have recalculated the decay rate of the electroweak vacuum in the SM using the corrected counting, 
and found it increases a little.
Even though the numerical difference compared to previous results is rather small, it is important to understand 
the conceptual issue with previous analyses and give a fully consistent picture of the meta-stability of the SM vacuum.
The argument on the degeneracy factor applies to any calculation of gauge determinants in $4D$.

{\bf \textsc{Acknowledgments.}}
We wish to thank So Chigusa for providing a sample dataset for our numerical check.
This work is supported by the Slovenian Research Agency under the research core funding No. P1-0035 and in 
part by the research grants N1-0253 and J1-4389.

\bibliographystyle{apsrev4-2}
\bibliography{references}

\clearpage

\begin{center}
  \large \bf End Matter
\end{center}

{\bf Previous vector field basis.}
In this Appendix we examine the basis for the vector field first introduced in Eq.~(4.44) of~\cite{Isidori:2001bm} and
then used in~\cite{Chigusa:2017dux, Chigusa:2018uuj, Andreassen:2017rzq}.
We show that the degeneracy factor of the transverse modes assumed in those works was incorrect.
We write their basis as
\begin{align} \label{eq:IRS_Amux}
\begin{split}
  A_\mu(x) = \sum_{j,m_A, m_B} & \left[ 
  A^{(B)}_{j m_A m_B}(\rho) B^{j m_A m_B}_\mu(\hat x)  \right.  \\ 
  & \ + A^{(L)}_{j m_A m_B}(\rho) L^{j m_A m_B}_\mu(\hat x)
    \\
  & \  + A^{(T1)}_{j m_A m_B}(\rho)T^{1; j m_A m_B}_\mu(\hat x)   \\
  &   \ \left. +  A^{(T2)}_{j m_A m_B}(\rho)T^{2;j m_A m_B}_\mu(\hat x) \right] \, .
\end{split}
\end{align}
Here, $A^{(B,L,T1,T2)}_{j m_A m_B}(\rho)$ are the radial parts, functions of the Euclidean radius $\rho$ only, and
\begin{align} \label{eq:IRS_orbital}
\begin{split}
  B^{j m_A m_B}_\mu(\hat x)	&= \hat x_\mu \, 		Y_{j m_A m_B}(\hat x) \, ,  \\
  L^{j m_A m_B}_\mu(\hat x)	&= \rho \partial_\mu \, 	Y_{j m_A m_B}(\hat x) \, , \\
   T^{i; j m_A m_B}_\mu(\hat x) &= i \epsilon_{\mu \nu \rho \sigma} V^{(i)}_\nu L_{\rho \sigma} Y_{j m_A m_B}(\hat x)\, ,
  \end{split}
\end{align}
with $i = 1,2$, are the angular basis elements, which are only functions of three angles in 4D.
The $Y_{j m_A m_B}$ are eigenfunctions of the $L^2 = L^{\mu\nu}L_{\mu\nu} / 2$ operator
with $L^2 Y_{j m_A m_B} = j(j + 2) Y_{j m_A m_B}$.
Here, the label $j$ is an integer, while $m_A$ and $m_B$ range between $-j/2$ and $j/2$ in integer steps, giving the multiplicity $(j+1)^2$.
$B$ is the `breathing mode' along the direction $\hat x_\mu$ and $L$ the longitudinal one, along the momentum
$\rho \partial_\mu$.
The $T^{1,2}$ are the transverse modes, with $V^{(1)}_\mu$ and $V^{(2)}_\mu$ being two arbitrary independent 
vectors, orthogonal to both $B$ and $L$. 
Finally, $\epsilon_{\mu \nu \rho \sigma}$ is the Levi-Civita antisymmetric tensor.

In general, one can construct different bases for vectors and the number of eigenstates of $L^2$ 
with a fixed eigenvalue has to be the same, i.e. basis independent.
The simplest way to count such states is to consider a basis made of four vectors of the form 
$\delta_{\mu i} Y_{j m_A m_B}$, with $i=1,2,3,4$.
Each of them inherits the $L^2$ eigenvalue of $j(j+2)$ and each carries the degeneracy of $(j+1)^2$,
totalling in $4(j+1)^2$ (for more details, see the last section of the Supplemental material).

Now let us consider the vector basis introduced in~\cite{Isidori:2001bm}, count the number of 
eigenstates of $L^2$ with a fixed eigenvalue of $j(j+2)$ and check whether they add up to $4(j+1)^2$.
The transverse modes $T^{i; j m_A m_B}_\mu$ themselves are already eigenfunctions of $L^2$ with the 
eigenvalue $j(j+2)$.
Although a priori their degeneracy is not obvious, Ref.~\cite{Isidori:2001bm} assumes that the two of them account
for $2(j+1)^2$ independent states.
On the other hand, the modes $B^{j m_A m_B}_\mu$ and $L^{j m_A m_B}_\mu$ are not eigenfunctions of 
$L^2$, but the following 
linear combinations, already derived in~\cite{Isidori:2001bm}, are:
\begin{align} \label{eq:f12j} 
\begin{split}
  f_1^{j m_A m_B} & \propto  \left( j + 1 \right) B^{(j + 1) m_A m_B}_\mu + L^{(j + 1) m_A m_B}_\mu \, ,
  \\
  f_2^{j m_A m_B} & \propto  \left( j + 1 \right) B^{(j - 1) m_A m_B}_\mu - L^{(j - 1) m_A m_B}_\mu \, .  
\end{split}  
\end{align}
Their eigenvalue of $L^2$ is $j(j+2)$ and the associated degeneracy factors are $(j+2)^2$ and $j^2$, respectively. 
Adding up all the degeneracies of the $f_{1, 2}$ and $T^{1, 2}$ basis, we have $(j+2)^2 + j^2 + 2(j+1)^2 = 4(j+1)^2 + 2$.
Comparing to the expected result, we got two too many.
What went wrong?
It is the degeneracy assumed for the transverse modes. 
The correct combined degeneracy should be $2j(j+2)$, which is two units less than $2(j+1)^2$.

The point is that the linear transformation in~\eqref{eq:IRS_orbital}, defining $T^{i; j m_A m_B}_\mu$
out of $Y_{j m_A m_B}$, is not injective and we lose degrees of freedom from the $(j+1)^2$ of the hyperspherical harmonics.
As a simple demonstration, let us consider the case with $j=1$.  
We take the following real combination of $Y_{1 m_A m_B}$,
\begin{equation}
Y_{1 \alpha} = \left\{ \frac{x}{\rho}, \frac{y}{\rho}, \frac{z}{\rho}, \frac{t}{\rho}  \right\} = \frac{x_\alpha}{\rho} 
= \hat x_\alpha \, ,
\end{equation}
which gives us four independent objects.
Then one set of transverse modes with $V_\nu^{(1)} = (1,0,0,0) = \delta_{\nu 1}$, is given by
\begin{align}
\begin{split}
  &T_\mu^{1;(\alpha)}  = \epsilon_{\mu\nu\rho\sigma} V_\nu^{(1)} x_\rho \partial_\sigma Y_{1 \alpha}
  \\
  & = \epsilon_{\mu\nu\rho\sigma} \delta_{\nu 1} x_\rho \left( \frac{\delta_{\sigma\alpha}}{\rho} - \frac{x_\sigma x_\alpha}{\rho^3}  \right)
  = \epsilon_{\mu 1 \rho\alpha} \hat x_\rho \, .
\end{split}
\end{align}
We see that $T_\mu^{1;(\alpha = 1)} = 0$ and a degree of freedom is lost.
The other transverse modes, taking $V_\nu^{(2)} = (0,1,0,0) = \delta_{\nu 2}$, are
$T_\mu^{2;(\alpha)} = \epsilon_{\mu 2 \rho\alpha} \hat x_\rho$,
so that $T_\mu^{2;(\alpha = 2)} = 0$.
It is then easy to check that, out of the remaining 6 transverse modes, only 5 are linearly independent, 
because $T_\mu^{1;(\alpha = 2)} = -T_\mu^{2;(\alpha = 1)}$.

We have thus shown that the degeneracy factor of the transverse modes defined in~\cite{Isidori:2001bm} is not
$2(j+1)^2$, as assumed in that work and in subsequent literature.
The functions $T^{1, 2}$ actually cover a space with dimension $2 j(j+2) - j$ for any $j>0$, as proven for
generic $j$ in the Supplemental Material, and exemplified above for $j=1$.
This is neither the na\"ive $2(j+1)^2$, nor the correct $2j(j+2)$. 
This demonstrates that $T^{1,2}$ from~\eqref{eq:IRS_orbital} are not linearly independent and do not form a complete 
basis for the transverse modes. 
Additional technical details of this issue are discussed in the Supplemental Material.

\clearpage
\onecolumngrid

\begin{center}
  \large \bf Supplemental Material
\end{center}

\section{Previous vector field basis}
In this section we examine the basis for the vector field first introduced in~\cite{Isidori:2001bm} and used 
in~\cite{Chigusa:2017dux, Chigusa:2018uuj, Andreassen:2017rzq}.
We write it as
\begin{align} \label{eq:IRS_Amu}
\begin{split}
  A_\mu &= \sum_{j = 0}^\infty \sum_{m_A, m_B} \left(
  A^{(B)}_{j m_A m_B}(\rho) B^{j m_A m_B}_\mu(\hat x) + A^{(L)}_{j m_A m_B}(\rho) L^{j m_A m_B}_\mu(\hat x)\right.
  \\
  & \quad \qquad \qquad \qquad \left. + A^{(T1)}_{j m_A m_B}(\rho)T^{1; j m_A m_B}_\mu(\hat x) + 
  A^{(T2)}_{j m_A m_B}(\rho)T^{2;j m_A m_B}_\mu(\hat x) \right) \, ,
\end{split}
\end{align}
where $A^{(B,L,T1,T2)}_{j m_A m_B}(\rho)$ are the radial parts, functions of the Euclidean radius $\rho$ only. 
Here, the labels $j, m_A$ and $m_B$ are exactly the labels $j, m$ and $m'$ introduced in~\cite{Isidori:2001bm}, 
which indicate the hyperspherical harmonics that the basis functions are constructed from.
The orbital parts do not depend on $\rho$ and are only functions of $\hat x$, or the three angles in 4D.
We separate them into $B$ modes, proportional to the `breathing' direction $\hat x_\mu$, and longitudinal-like 
$L$ modes along the momentum $\rho \partial_\mu$:
\begin{align} \label{eq:IRS_BL}
  B^{j m_A m_B}_\mu(\hat x)	&= \hat x_\mu \, 		Y_{j m_A m_B}(\hat x) \, ,
  &
  L^{j m_A m_B}_\mu(\hat x)	&= \rho \partial_\mu \, 	Y_{j m_A m_B}(\hat x) \, .
\end{align}
Orthogonally to these two, the authors of \cite{Isidori:2001bm} defined the transverse modes,
\begin{align} \label{eq:IRS_T12}
  T^{i; j m_A m_B}_\mu(\hat x) &= i \epsilon_{\mu \nu \rho \sigma} V^{(i)}_\nu L_{\rho \sigma} Y_{j m_A m_B}(\hat x) \, , 
  &  i &= 1,2 \, ,
\end{align}
where $V^{(1)}_\mu$ and $V^{(2)}_\mu$ are two arbitrary independent vectors.
The construction relies on the hyperspherical harmonics that we label as $Y_{j m_A m_B}$. 
They are eigenfunctions of the $L^2$ operator, $L^2 Y_{j m_A m_B} = j(j + 2) Y_{j m_A m_B}$, where the label $j$ is
an integer, while $m_A$ and $m_B$ range between $-j/2$ and $j/2$ in integer steps, giving the multiplicity $(j+1)^2$. 
Na\"ively, it looks like each set of functions, $B, L, T^1, T^2$, spans a space with dimension $(j+1)^2$ for any $j>0$, 
which is what was assumed in~\cite{Isidori:2001bm, Chigusa:2017dux, Chigusa:2018uuj, Andreassen:2017rzq}. 
We will show that this is not the case.

In particular, the transverse modes $T^{1;j}$ and  $T^{2;j}$ only cover a space with dimension $2j(j+2) - j$ for any $j>0$, 
which is definitely not equal to the na\"ive $2(j+1)^2$. 
As we will show later, the correct degeneracy is $2j(j+2)$, so the basis of \cite{Isidori:2001bm} misses $j$ elements, 
meaning that it is both dependent and incomplete.
Let us start with the $B$ and $L$ modes, which are orthogonal to each other.
The $j = 0$ mode is constructed from $Y_{000}(\hat x) = 1/\sqrt{4 \pi}$.
Therefore, only $B_\mu^{000}$ is non zero, while $L_\mu^{000}$ and $T^{i;000}_\mu$ vanish because they contain
a derivative hitting the constant $Y_{000}$.
Also, $B_\mu^{000}$ is proportional to $V^{00}_{l=1;\mu}$ [see ({\color{nicered}9})], which corresponds 
to the singlet state.
Next, consider $j \geq 1$. 
Both $B^{j m_A m_B}_\mu$ and $L^{j m_A m_B}_\mu$ are non-zero for all $m_A$ and $m_B$.
Note that the $\hat x^\mu$ and $\partial_\mu$ operators commute with $J_{\mu \nu}$.
This implies that $B^{j m_A m_B}_\mu$ and $L^{j m_A m_B}_\mu$ have the total angular momentum quantum 
numbers $j_A = j_B = j/2$, and are orthogonal to each other.
They are eigenfunctions of $J^2$ with eigenvalue $j(j+2)$, and have the same degeneracy factor as the 
hyperspherical harmonics, $(j+1)^2$.

Another way of understanding the degeneracy is to calculate the number of independent functions from the rank 
of the Gram matrix formed from the basis functions.
We have
\begin{align}
  d_j^B &	= \rank \left(\sum_\mu \int{\rm d} \Omega (B^{j m_A m_B}_\mu)^* B^{j m'_A m'_B}_\mu \right)
  		= \rank \left(\delta_{m_Am'_A}\delta_{m_Bm'_B}\right) = \left( j + 1 \right)^2\,,
\end{align}
where we plugged in the definition of $B^{j m_A m_B}_\mu$ and used the fact that $\sum_\mu \hat x_\mu^2 = 1$. 
This reduces the integral to the usual statement of spherical harmonic orthonormality after integrating over $\int{\rm d}\Omega$.
We also merged $m_A$ and $m_B$ into a single index and the Gram matrix becomes a $(j + 1)^2 \times (j + 1)^2$ matrix.
A similar calculation goes through for the longitudinal-like modes, where we integrate by parts and express the 
Cartesian second derivative in the spherical basis as $\Delta_l$.
The radial derivatives vanish when we act on $\partial_\rho^n ((B, L)(\hat x)) = 0$ and only the orbital momentum 
operator $L^2/\rho^2$ part remains, such that
\begin{align} \label{eq:dLIRS}
  d_j^L &	= \rank \left(\sum_\mu \int{\rm d}\Omega (L^{j m_A m_B}_\mu)^* L^{j m'_A m'_B}_\mu \right) 
  		= \rank \left(j (j + 2) \delta_{m_A m'_A}\delta_{m_B m'_B} \right) =
  \begin{cases}
    0 \, , & j = 0 \, ,
    \\
    (j+1)^2 \, , & j > 0 \, .
  \end{cases}
\end{align}
The $L^2$ eigenvalue of $Y_{j m_A m_B}$ is simply $j (j + 2)$ and is just a number that does not affect the rank for 
$j > 0$; it only takes care of removing the $j = 0$ mode.
As expected, the Gram matrix for generic $j > 0$ is full-rank, except for $L_\mu^{000}$ that vanishes.

We have recovered the result of the group theoretical argument in the main text.
It follows that the modes $B^{j m_A m_B}_\mu$ and $L^{j m_A m_B}_\mu$
cover exactly the same space spanned by our ${\cal D}_j^\pm$ basis, which is a $2(j + 1)^2$ dimensional space.
Unlike $V^{j_A j_B}_{l;\mu}$, the $B^{j m_A m_B}_\mu$ and $L^{j m_A m_B}_\mu$ are still not eigenstates 
of $L^2$ and are diagonalized by taking linear combinations,
 \begin{align}
  V^{\frac{j}{2} \frac{j}{2}m_Am_B}_{l=j+1;\mu} & \propto \left( j + 2 \right) \, B^{j m_A m_B}_\mu - L^{j m_A m_B}_\mu \, ,
  &
  V^{\frac{j}{2} \frac{j}{2}m_Am_B}_{l=j-1;\mu} 	& \propto j \, B^{j m_A m_B}_\mu + L^{j m_A m_B}_\mu \, ,
\end{align}
where the first function corresponds to the ${\cal D}_j^+$ multiplet and the second to ${\cal D}_j^-$.
The degeneracy factor for these modes was treated correctly in previous literature. 
Let us then turn to the transverse modes, for which the situation is more involved.

\section{Transverse gauge modes}
First note that the $T^{i; j m_A m_B}_\mu$ of \eqref{eq:IRS_Amu} are not eigenstates of $J_{A3, B3}$, the third component 
of the $SU(2)_A$ and $SU(2)_B$ total angular momentum.
Hence, the labels $m_A$ and $m_B$ do not correspond to the eigenvalues of $J_{A3}$ and $J_{B3}$, they are 
just inherited from the labels of spherical harmonics. Thus having different $m_A$ and $m_B$ does not ensure 
that the functions are independent.
The $J^2$ eigenvalue for $T^{i; j m_A m_B}_\mu$ is given by $(j+1)^2$, therefore they have to be related to 
the ${\cal T}^\pm_j$ modes. 
However, we know from the discussion in the main text that there exist only 
$2 j (j + 2)$ independent basis functions for ${\cal T}^\pm_j$.
Let us then analyze the space spanned by $T^{1,2; j m_A m_B}_\mu$.

{\bf Computing the degeneracy of $T^1$.}
We invoke again the rank of the Gram matrix.
First we need some preliminaries.
Without loss of generality we take $V^{(1)}_\mu = \delta_{\mu4}$ and $V^{(2)}_\mu = \delta_{\mu3}$ as the 
two independent vectors in~\eqref{eq:IRS_T12}.
Then we define the $SU(2)_A$ and $SU(2)_B$ orbital angular momentum operators as
\begin{align}
  L_{A i} &= \frac{1}{4} \left( \epsilon_{ijk} L_{jk} + 2L_{4i} \right) \, ,
  &
  L_{B i} &= \frac{1}{4} \left( \epsilon_{ijk} L_{jk} - 2L_{4i} \right) \, .
\end{align}
They are normalized such that they satisfy the canonical $SU(2)_{A, B}$ algebras 
$[ L_{A i}, L_{A j} ] = i \epsilon_{ijk} L_{A k}$, 
$[L_{B i}, L_{B j}] = i \epsilon_{ijk} L_{B k}$ and commute with one another $[L_{A i}, L_{B j} ] = 0$.
Then we have
\begin{align} \label{eq:dT1IRS}
  d_j^{(T1)} &	= \rank \left( \sum_\mu \int{\rm d} \Omega \, (T^{1; j m_A m_B}_\mu)^* T^{1; j m'_A m'_B}_\mu \right)
			= \rank \left( \int{\rm d} \Omega \, Y^*_{j m_A m_B} 4 \sum_i \left( L_{A i} + L_{B i} \right)^2 Y_{j m'_A m'_B} \right) \,.
\end{align}

Let us evaluate the action of the $L_{A i} + L_{B i} = L_{C i}$ operator on spherical harmonics 
in~\eqref{eq:dT1IRS}.
Because $A$ and $B$ sectors commute, the $C$ operators themselves form an $SU(2)_C$ algebra that 
represents a simultaneous rotation in both $SU(2)_{A,B}$ spaces.
Its states can be labeled by $l_C$ that ranges from $|l_A - l_B| \leq l_C \leq l_A + l_B$.
The hyperspherical harmonics $Y_{j m_A m_B}$ correspond to $(l_A, l_B) = (j/2, j/2)$.
Because $l_A = l_B$, the minimal value of $l_C$ is always $0$ for each fixed $j$ in~\eqref{eq:dT1IRS}.
The eigenvalue of the Casimir operator $L_C^2 = \sum_i \left( L_{A i} + L_{B i} \right)^2$ is in general given 
by $l_C(l_C+1)$ (as for the canonical $SU(2)_C$) and therefore vanishes for $l_C = 0$.
This means that there will always be a single zero eigenvalue in the Gram matrix of $T^1$, the rest are non-zero.
Without the zero, the matrix would be full rank with the degeneracy of $(j+1)^2$, because none
of the states would be lost in the last equality of~\eqref{eq:dT1IRS}.
However, the zero is present for each $j$ (this is in contrast to~\eqref{eq:dLIRS}, where the zero is there 
only for $j=0$) and therefore reduces $d_j^{(T1)}$ by 1 and we have
\begin{equation} \label{eq:GramT1}
  d_j^{(T1)} = \left( j + 1 \right)^2 - 1 = j \left( j + 2 \right) \, .
\end{equation}

The vanishing linear combination among the $T^1$ modes, whose existence is guaranteed by the previous 
arguments and which corresponds to the singlet of $L_C$, is given explicitly by
\begin{equation} \label{eq:singlet}
  0 = \sum_{m = -j/2}^{j/2}(-1)^m \, T^{1;jm(-m)}_\mu \, .
\end{equation}
Had we started with a different choice of $V^{(i)}_\mu$, by rotational symmetry of the problem we would 
have obtained the same result, therefore $ d_j^{(T2)} = j \left(j + 2 \right)$.
Furthermore, in analogy to~\eqref{eq:singlet}, there is a vanishing linear combination among the $T^2$ 
modes, given explicitly as $0 = \sum_{m = -j/2}^{j/2} T^{2;jm(-m)}_\mu$.

With the calculated $d_j^{(T1,T2)}$ it looks like we got exactly what was expected from group theoretical grounds.
It appears that the calculation of the transverse mode determinant, using the basis proposed in~\cite{Isidori:2001bm},
would be perfectly fine, if only the correct degeneracy factor were used.
However, this is not the end of the story, as it turns out that $T^1$ and $T^2$ have an overlapping space, 
as we are going to show.

{\bf Additional overlaps among $T^{1,2}$ transverse modes.}
We want to prove here that the transverse modes $T^{1;j}$ and $T^{2;j}$, defined in~\eqref{eq:IRS_T12}, do not span the 
whole $2 j (j + 2)$ dimensional space of transverse fluctuations.
Specifically, at each $j$ we find that precisely $j$ modes are linearly dependent, such that
\begin{equation}\label{eq:T12spacedim}
  d_j^{(T1 \cup T2)} = d_j^{(T1)} + d_j^{(T2)} - j=j(2j+3) \, ,
\end{equation}
where we called $d_j^{(T1 \cup T2)}$ the dimension of the space generated by \eqref{eq:IRS_T12} at a given $j$. 

To better understand the nature of the problem, it is useful to split the $T^{1,2}$ modes into those having 
$m_A + m_B \neq 0$ and those having $m_A + m_B = 0$. 
The first class is made of $j(j + 1)$ $T^1$ modes and $j(j + 1)$ $T^2$ modes, while the second class is made of 
$j$ $T^1$ modes and $j$ $T^2$ modes.
The reason why there are only $j$ modes in the second class, and not $j+1$, is due to~\eqref{eq:singlet} and its 
homologous for $T^2$ modes.
Notice in fact that the vanishing linear combinations only involve modes of the form $T^{i;jm(-m)}$.
Due to the Gram determinant discussion, resulted in \eqref{eq:GramT1}, we also know that there is no other 
vanishing linear combination within the set of $T^1$ modes, and in particular all those $j(j+1)$ modes with 
$m_A+m_B\neq 0$ are linearly independent.
The same holds for $T^2$.
However we cannot say much, for the moment, about possible linear relations among the combined set 
comprising both $T^1$ and $T^2$ modes.
This is what we are going to investigate next.

Firstly, we are going to prove that all $T^1$ and $T^2$ modes belonging to the first class with $m_A+m_B\neq 0$
are linearly independent.
With a little bit of algebra it can be shown that the $T^{1,2}$ modes take on the following form
\begin{align}\label{eq:T12indep}
  T^{1; j m_A m_B}_\mu & \propto 
  \begin{pmatrix}
    * \\
    * \\
    m_A + m_B  \\
    0
  \end{pmatrix} Y_{j m_A m_B} \,,
  &
  T^{2; j m_A m_B}_\mu & \propto 
  \begin{pmatrix}
    * \\
    * \\
    0 \\
    m_A + m_B  \\
  \end{pmatrix} Y_{j m_A m_B} \,,
\end{align}
where $*$ indicates some linear combinations of $L_{\mu\nu}$.
Thanks to the linear independence of the spherical harmonics and the fact that $m_A + m_B\neq 0$, it follows 
that all modes belonging to the first class are independent, which means that they span altogether a 
$2j(j+1)$-dimensional space.
For what concerns modes with $m_A + m_B = 0$, which have only zeroes in the last two components, by 
inspection of~\eqref{eq:T12indep} we can also conclude that they generate a subspace which is linearly 
independent from the one generated by the modes with $m_A + m_B\neq 0$. 
What is the dimension of this subspace is our next question.

With a little bit of algebra, we find
\begin{align}
  T^{1; j m (-m)}_\mu &=  
  \begin{pmatrix}
    i(L_{A +} + L_{A -} + L_{B+} + L_{B-}) \\
    L_{A +} - L_{A -} + L_{B+} - L_{B-}   \\
    0 \\
    0
  \end{pmatrix} Y_{j m (-m)} \,,
  \\
  T^{2; j m (-m)}_\mu &= 
  \begin{pmatrix}
    L_{A +} - L_{A -} - L_{B+} + L_{B-} \\
    i(-L_{A +} - L_{A -} + L_{B+} + L_{B-}) \\
    0 \\
    0
  \end{pmatrix} Y_{jm(-m)} \,,
\end{align}
where the orbital ladder operators are defined in the usual way as $L_{A \pm} = L_{A 1} \pm i L_{A 2}$ and 
$L_{B \pm} = L_{B 1} \pm i L_{B 2}$.
Working out the action of the ladder operators on the spherical harmonics, one finds that the nonzero components 
of $T^{1,2}$ have the following form
\begin{align}\label{eq:T12m-m}
  T^{1; j m(-m)}_\mu & = 
  \begin{pmatrix}
    i(P_m+P_{m-1})  \\
    Q_m+Q_{m-1}
  \end{pmatrix} \, ,
  &
  T^{2; j m(-m)}_\mu & =
  \begin{pmatrix}
    P_m-P_{m-1}  \\
    -i(Q_m-Q_{m-1})
  \end{pmatrix} \, ,
\end{align}
where, at a given $j$,
\begin{align}
  P_m,Q_m &= \sqrt{\left(\frac{j}{2}+m+1\right)\left(\frac{j}{2}-m\right)} \left(Y_{j(m+1)(-m)}\,\pm\, Y_{jm(-m-1)}\right) \, .
\end{align}
By inspection of the second expression of \eqref{eq:T12m-m}, knowing that $P_{-\frac{j}{2}-1},Q_{-\frac{j}{2}-1}=0$, 
we find that
\begin{equation}\label{eq:T2sum}
i\sum_{\sigma=-\frac{j}{2}}^m T^{2; j \sigma(-\sigma)}_\mu=\begin{pmatrix}
    iP_m  \\
    Q_m
  \end{pmatrix} \, ,
\end{equation}
(a completely analogous expression can be obtained by taking an alternate sum over $T^1$ modes).
Eq.~\eqref{eq:T2sum} allows to derive the following set of $j + 1$ identities, expressing $T^{1;j m (-m)}_\mu$ 
modes as a linear combination of $T^{2;j m (-m)}_\mu$ modes
\begin{equation} \label{eq:overlap}
    T^{1;j m (-m)}_\mu = i \sum_{\sigma = -\frac{j}{2}}^{m}T^{2; j \sigma (-\sigma)}_\mu + i \sum_{\sigma = -\frac{j}{2}}^{m-1}T^{2;j \sigma(-\sigma)}_\mu \, .
\end{equation}
Let us explain why this equality allows to conclude that Eq.~\eqref{eq:T12spacedim} holds.
We know for sure from the Gram determinant discussion that there are exactly $j$ independent $T^2$ 
modes of the form $T^{2;jm(-m)}$, which are also independent from all modes with $m_A+m_B\neq 0$, as 
we just showed. 
Eq.~\eqref{eq:overlap} shows that all $T^1$ modes of the form $T^{1;jm(-m)}$ are linearly dependent 
(on the $T^{2;jm(-m)}$ modes). 
Even though we declare $j+1$ linear identities, exactly one of them is redundant, because of~\eqref{eq:singlet}. 

Summarizing, we have found that there are $2j(j+1)$ independent modes having $m_A + m_B\neq 0$, and 
that out of the initial $2j$ modes having $m_A + m_B = 0$, which a span an independent space on their own, 
only $j$ are independent (say the $T^2$ modes).
All in all we have $d_j^{(T1\cup T2)} = 2 j \left( j + 1 \right) + j = j \left( 2 j + 3 \right)$, as we wanted to show.

Our proof was built on making the choice $V^{(1)} = \delta_{\mu 4}, V^{(2)} = \delta_{\mu 3}$.
These two directions play a special role: the first in defining ``boosts'', the second in picking up a 
quantization axis.
Rotational invariance of the problem guarantees that any choice of two orthogonal vectors would lead 
to the same conclusion.
Indeed, any two orthogonal vectors can be rotated to $\delta_{\mu4}$ and $\delta_{\mu 3}$.

{\bf Completing the transverse basis.}
The missing $j$ transverse modes can be chosen as
\begin{align}\label{eq:T3}
  &T^{3; j (m-1)(-m)}_\mu & &\text{with}  & V_{\nu}^{(3)} &= \delta_{\nu1} + i \delta_{\nu2} \, .
\end{align}
We see indeed that the fourth component of $T^{3; j (m-1)(-m)}_\mu$ is
\begin{align}
  T^{3;j (m - 1)(-m)}_{\mu=4} = 
  -2 i \sqrt{\left(\frac{j}{2} + m \right) \left(\frac{j}{2} + 1 \right)} Y_{j m (-m)} - 
   2 i \sqrt{\left(\frac{j}{2} - m + 1 \right) \left(\frac{j}{2} + m\right)} Y_{j (m-1)(-m+1)} \, ,
\end{align}
which is independent from $T^{1,2}_{\mu=4}$ because it contains the missing $Y_{jm(-m)}$. 
Independence of the fourth component implies independence of the $j$ vectors in \eqref{eq:T3} from 
the previously considered vectors.
Now, $T^{2;j m_Am_B}$ with arbitrary $m_{A,B}$ modulo the singlet removal (in total $j(j+2)$ modes), 
$T^{1;j m_Am_B}$ with $m_A+m_B\neq 0$ (in total $j(j+1)$ modes) and $T^{3;j (m - 1)(-m)}$ 
(in total $j$ modes) are independent and correctly span the whole $2 j(j + 2)$ dimensional space of 
transverse fluctuations. 

\section{Canonical basis of transverse modes}
An alternative to the basis given in the previous section, one that elaborates on the ansatz of~\cite{Isidori:2001bm}, 
Eq.~\eqref{eq:IRS_T12}, is given by the following expression
\begin{equation}\label{eq:Taltern}
  T^{\pm; j m_A m_B}_\mu(\hat x)  =
  \sum_{m_{l_A}, m_{l_B}}  \epsilon_{\mu \nu \rho \sigma}
\left({\bf V}^\pm_\nu\right)^{m_A \ m_B}_{m_{l_A} m_{l_B}}
  L_{\rho\sigma} Y_{j m_{l_A} m_{l_B}}(\hat x) \, ,
\end{equation}
where the matrix-valued 4-vector ${\bf V}_\nu^\pm$, converting the $(j/2,j/2)$ indices of the spherical 
harmonics to $((j\pm 1)/2,(j\mp 1)/2)$ indices (the ones appropriate for the transverse modes, as discussed 
in the main text), is given in terms of Clebsch-Gordan coefficients as
\begin{equation}
  \left({\bf V}^\pm_\nu\right)^{m_A \ m_B}_{m_{l_A} m_{l_B}}= \sum_{m_{s_A}, m_{s_B}} 
  C^{\frac{j \pm 1}{2} m_A}_{\frac{j}{2} m_{l_A} \frac{1}{2} m_{s_A}} 
  C^{\frac{j \mp 1}{2} m_B}_{\frac{j}{2} m_{l_B} \frac{1}{2} m_{s_B}} 
  \left( \tilde \sigma_\nu \right)_{m_{s_A} m_{s_B}} \, ,
\end{equation}
with $\tilde \sigma_\nu$ as in ({\color{nicered}9}). 
Eq.~\eqref{eq:Taltern} is nothing but a linear combination of objects of the form of \eqref{eq:IRS_T12}. 
The virtue of this specific linear combination is that the $m_{A,B}$ indices now \emph{do} correspond to the 
eigenvalues of the $J_{A3}$ and $J_{B3}$ operators, making \eqref{eq:Taltern} a canonical multiplet of 
$SO(4)\simeq SU(2)\times SU(2)$. 
Moreover, the ansatz of~\cite{Isidori:2001bm} makes transversality of the modes manifest.

Let us comment on the mechanics behind \eqref{eq:Taltern}. 
The main observation is that, under a rotation acting on the vector field $T^{\pm;j}_\mu(\hat{x})$, it can be shown 
that the combination 
$\epsilon_{\mu \nu \rho \sigma}( \tilde \sigma_\nu )_{m_{s_A} m_{s_B}}L_{\rho\sigma} Y_{j m_{l_A} m_{l_B}}(\hat x)$ 
transforms as the tensor product $(j/2,j/2)\otimes(1/2,1/2)$, where the first and second factor act respectively on 
the indices $m_{l_A} ,m_{l_B}$ and $m_{s_A} ,m_{s_B}$. 
In order to project on the $((j\pm 1)/2,(j\mp 1)/2)$ subrepresentations, the ones characterizing transverse modes, 
we contract with $C^{\frac{j \pm 1}{2}}C^{\frac{j \mp 1}{2}}$. 
This is completely analogous to the standard procedure with the rotation group $SO(3)$, with the difference here that 
there are two factors of $SO(3)\simeq SU(2)$, so we need to multiply by two Clebsch-Gordan coefficients. 
As a cross check of the consistency of the method, it can be verified that contracting with 
$C^{\frac{j + 1}{2}}C^{\frac{j + 1}{2}}$ or with $C^{\frac{j - 1}{2}}C^{\frac{j - 1}{2}}$ gives zero, because diagonal modes 
are already projected out by the 
$\epsilon_{\mu\nu\rho\sigma}$. 
The modes of~\eqref{eq:Taltern} are proportional to the ${\cal T}_j^{\pm}$ modes presented in the main text.

\section{Counting in orbital modes}
Another possible basis for the vector field is
\begin{equation} \label{vectorbital}
A_\mu(x) = \sum_{l, m_{l_A}, m_{l_B}} \left( \delta_{\mu 1} A_{l m_{l_A} m_{l_B}}^{(1)}(\rho)  +
\delta_{\mu 2} A_{l m_{l_A} m_{l_B}}^{(2)}(\rho)  +
\delta_{\mu 3} A_{l m_{l_A} m_{l_B}}^{(3)}(\rho) + 
\delta_{\mu 4} A_{l m_{l_A} m_{l_B}}^{(4)}(\rho)  \right) Y_{l m_{l_A} m_{l_B}} (\hat x) \, .
\end{equation}
Here we use the label $l$ to emphasize that the hyperspherical harmonics are eigenfunctions of $L^2$, 
the Casimir of the orbital angular momentum operator. 
The form \eqref{vectorbital} makes it obvious that the vector can be seen as a collection of four scalar 
degrees of freedom. 
As such, for each $l$ the vector has degeneracy of $4(l+1)^2$.
Let us count the degeneracies in terms of the orbital angular momentum quantum number $l$ for the 
basis we introduced in the main text, and for the $B,L,T^i$ basis we discussed just above.

We saw that the transverse modes in either basis, ${\cal T}_j^\pm$ or $T^{i ;j}_\mu$, have $l = j$.
Their degeneracy is $2l(l+2) = 2j(j+2)$.
The other two modes with $l = j$ are
\begin{align}
    V^{\frac{j+1}{2} \frac{j+1}{2}}_{l=j;\mu} \propto 
    \left( l + 1 \right) B^{(l + 1) m_A m_B}_\mu + L^{(l + 1) m_A m_B}_\mu \in \left( \frac{l + 1}{2}, \frac{l + 1}{2}\right)_{l=j} \,,
    \\
    V^{\frac{j-1}{2} \frac{j-1}{2}}_{l=j;\mu} \propto
    \left( l + 1 \right) B^{(l - 1) m_A m_B}_\mu - L^{(l - 1) m_A m_B}_\mu    \in \left( \frac{l - 1}{2}, \frac{l - 1}{2}\right)_{l=j} \,,
\end{align}
which have $(l + 2)^2$ and $l^2$ degeneracies, respectively. Summing all of them up, we have 
\begin{equation}
  \left( l + 2 \right)^2 + l^2 + 2 l \left( l + 2 \right) = 4 \left( l + 1 \right)^2 \, ,
\end{equation}
which is consistent with that of four scalars, as expected.

The basis~\eqref{vectorbital} looks much simpler than the one we adopted in the main text or the 
one introduced in~\cite{Isidori:2001bm}. 
However, its modes are not eigenfunctions of the total angular momentum operators ($J_A^2, J_B^2, J_{A3}, J_{B3}$). 
As we stressed in the main text, the fluctuation operator~({\color{nicered}3}), which acts on the 
$(A_\mu,\varphi)$ basis, always commutes with $J_{\mu \nu}$, but in general not with $L_{\mu \nu}$ and $S_{\mu \nu}$ separately. 
In order to take advantage of symmetries to diagonalize the fluctuation operator, it is therefore much more convenient
to use a basis for $A_\mu$ like the one we chose in this work, or the one chosen in~\cite{Isidori:2001bm}.

\end{document}

%% file: MultipoleScheme.tikz
\begin{tikzpicture}[line width=1.2 pt, scale=.7, baseline=(current bounding box.center)]
    
    	\draw[-stealth] (0,0) to  (0,5.5) ;
    	\draw[-stealth] (0,0) to  (5.5,0) ;
    
    	\foreach \x in {0,1,2}
		{\node at (2*\x,-.5) {\footnotesize $\x$} ;
		\node at (-.7,2*\x) {\footnotesize $\x$} ;}
	\foreach \x in {1,3,5}
		{\node at (\x,-.5) {\footnotesize $\frac{\x}{2}$} ;
		\node at (-.7,\x) {\footnotesize $\frac{\x}{2}$} ;}
		
	\foreach \x in {1,2,3,5} \foreach \y in {1,2,3,4,5}
  		\filldraw[line width=.1] (\x,\y) circle (1pt) ;
	\foreach \y in {1,2,4,5,6}
  		\filldraw[line width=.1] (4,\y) circle (1pt) ;
	\filldraw[line width=.1] (2,0) circle (1pt) ; 
	\filldraw[line width=.1] (0,2) circle (1pt) ;
	\filldraw[line width=.1] (6,4) circle (1pt) ; 
	\filldraw[line width=.1] (0,0) circle (1pt) ;
		
	\foreach \x in {0,1,2,3,4,5} 	\draw[line width =.4] (\x,\x) circle (5pt) ;
	\foreach \x in {1,2,4,5}	\draw[line width =.4] (\x+1,\x-1) circle (5pt);
	\foreach \x in {1,2,4,5}	\draw[line width =.4] (\x-1,\x+1) circle (5pt) ;
	\draw[blue,line width =.6] (4,2) circle (5pt) ;
	\draw[blue,line width =.6] (2,4) circle (5pt) ;
	\foreach \x in {1,3,5} 	\draw[line width =.4] (\x,\x) circle (8.8pt) ;
	\foreach \x in {2,4} 	\draw[blue,line width =.6] (\x,\x) circle (8.8pt) ;
		    
    	\node at (0,6) {$j_B$};
   	\node at (6,0) {$j_A$};
		
   	\draw[-stealth,blue,line width=.7] (3.3,3.3) to  (3.7,3.7) ;
    	\draw[-stealth,blue,line width=.7] (2.7,2.7) to  (2.3,2.3) ;
    	\draw[-stealth,blue,line width=.7] (2.7,3.3) to  (2.2,3.8) ;
    	\draw[-stealth,blue,line width=.7] (3.3,2.7) to  (3.8,2.2) ;

    	\node at (2,1.4) {\tiny $l=1,\textcolor{blue}{3}$} ;
    	\node at (4,1.55) {\tiny $l=\textcolor{blue}{3}$} ;
    	\node at (4,4.6) {\tiny $l=\textcolor{blue}{3},5$} ;
    	\node at (2,4.5) {\tiny $l=\textcolor{blue}{3}$} ;
    	\node at (4,3) {\tiny $l=2,4$} ;
 
\end{tikzpicture}